\documentclass[aps,prd,preprint,nofootinbib]{revtex4-1}

\usepackage{amsmath,amssymb,graphicx}
\usepackage{tabularx}
\newcolumntype{C}[1]{>{\centering\arraybackslash}p{#1}}
\usepackage{hyperref}

\usepackage{xcolor}

\newcommand{\be}{\begin{equation}}
\newcommand{\ee}{\end{equation}}
\newcommand{\bea}{\begin{eqnarray}}
\newcommand{\eea}{\end{eqnarray}}

\begin{document}
\author{Alessandro Casalino}
\email{alessandro.casalino@unitn.it}
\author{Aimeric Coll\'eaux}
\email{aimeric.colleaux@unitn.it}
\author{Massimiliano Rinaldi}
\email{massimiliano.rinaldi@unitn.it}
\author{Silvia Vicentini}
\email{silvia.vicentini@unitn.it}
\affiliation{Dipartimento di Fisica, Universit\`{a} di Trento,\\Via Sommarive 14, I-38123 Povo (TN), Italy}
\affiliation{Trento Institute for Fundamental Physics and Applications (TIFPA)-INFN,\\Via Sommarive 14, I-38123 Povo (TN), Italy}
\title{Regularized Lovelock gravity}
\begin{abstract}
A four-dimensional regularization of Lovelock-Lanczos gravity up to an arbitrary curvature order is considered. We show that Lovelock-Lanczos terms can provide a non-trivial contribution to the Einstein field equations in four dimensions, for spherically symmetric and Friedmann-Lema\^{i}tre-Robertson-Walker spacetimes, as well as at first order in perturbation theory around (anti) de Sitter vacua. We will discuss the cosmological and black hole solutions arising from these theories, focusing on the presence of attractors and their stability. Although curvature singularities persist for any finite number of Lovelock terms, it is shown that they disappear in the non-perturbative limit of a theory with a unique vacuum.
\end{abstract}

\allowdisplaybreaks[1]

\maketitle

\section{Introduction}

The Lovelock theorem \cite{Lovelock:1971vz} states that the most general (purely metric) gravitational theory leading to second order field equations in $d$-dimensions, for any class of metric fields, is given by the Lovelock-Lanczos action. In four dimensions and for generic values of the coupling constants, the theory reduces to General Relativity (GR) with a cosmological constant \cite{Lovelock:1972vz}, because the field equations associated with higher order Lovelock scalars are identically vanishing in this case.

However, let us consider for instance the Einstein tensor with a non-generic coupling containing a dimensional pole $G_{\mu\nu}/(d-2)$. Clearly, this expression is indeterminate for $d\to 2$ as the Einstein tensor is identically vanishing in this case. Therefore, such theory is not automatically contained within the assumptions of the Lovelock theorem, as it lacks a prescription on how to take the limit. In order to bypass that theorem, such prescription could yield regularized theories with a reduced notion of covariance, with non-conserved covariant field equations, with additional degrees of freedom, or with second order field equations for specific classes of geometries only, as it is usually the case in Quasi-Topological \cite{Quasi-Top1, Quasi-Top2} and Non-polynomial gravity theories \cite{Deser, LorenzoSergio, aim}.  

 In fact, the regularization of two-dimensional General Relativity has been shown to be well-defined in \cite{GRD=2} by Mann and Ross, resulting in a scalar-tensor theory with second order field equations. To the best of our knowledge, it is the oldest reference to such a dimensional regularization of a Lovelock-Lanczos scalar. 

A similar argument was used for the first time in the four-dimensional case by Tomozawa \cite{Tomozawa:2011gp}, who considered the following Einstein-Gauss-Bonnet theory
\begin{equation}
S = \int d^d x \sqrt{-g} \left( R + \frac{\alpha}{d-4} \mathcal{G} \right)\,,
\end{equation}
where $\mathcal{G} \equiv R^2 - 4 R^{\mu \nu} R_{\mu \nu} + R^{\mu \nu \rho \sigma} R_{\mu \nu \rho \sigma}$. He showed that when the corresponding field equations are specifically evaluated for a static spherically symmetric ansatz, then it becomes possible to smoothly take the limit $d=4$ without cancelling the contribution coming from the Gauss-Bonnet term. Thus, this constitutes what can be called a ``minisuperspace regularization" of this sector of Gauss-Bonnet gravity. 

The resulting model admits a black hole solution with a repulsive gravitational force near the time-like singularity $r\to 0$. Its geodesic structure as well as its quasi-normal modes and stability have been investigated recently in respectively \cite{geodesics} and \cite{QNM}. Moreover, the charged case was investigated in \cite{conformal} and more recently in \cite{charged}, while the AdS black holes and their thermodynamics were found and discussed in \cite{ConfThermo, thermod}. Finally, a rotating generalization has been studied in \cite{Rotating,Rotating2} using the Newman-Janis algorithm.

Using the same model, Cognola et al. \cite{Cognola:2013fva} showed that the entropy of this black hole is logarithmic in its area, which motivated an interpretation of these terms as effective quantum corrections, because it is well-known that such a behaviour of the entropy typically arises when quantum effects are taken into account, see for example \cite{Carlip,PerezNoui}. Moreover, they also established that the flat Friedmann-Lema\^{i}tre-Robertson-Walker (FLRW) sector of this theory is well-defined when taking the limit $d\to 4$. 

More recently,  Glavan and  Lin further extended the applicability of this "minisuperspace regularization" to first order in perturbation theory around (A)dS$_4$ vacua, showing that it contains the degrees of freedom of a massless graviton, as in GR \cite{Glavan:2019inb}. However, these results have been criticised on the basis that the limiting procedure is not well-defined for arbitrary metrics, see e.g. \cite{ConfGB2,Gurses:2020ofy,BianchiIGB2}.
Finally, one of us showed that non-singular black hole and cosmological solutions can be found considering non-perturbative curvature corrections, i.e. the whole Lovelock-Lanczos series  (see Chap. V, Sec. E.2. of \cite{aim}). 

Throughout the literature on black holes, FLRW cosmology and perturbations in the context of higher dimensional Lovelock-Lovelock gravity, this specific normalization of the coupling constants has been very common to avoid the presence of dimensional factors in the solutions, see for example \cite{Deruelle:1989fj, Maeda, Zanelli}. Therefore many results that were thought to be purely higher dimensional might be applicable as well in 4D.

Although this ``minisuperspace regularization" is not unique \cite{BianchiIGB2,ConfGB2} and cannot be extended to arbitrary metric fields \cite{TekinGB1,TekinGB2,Amplitudes}, this approach is interesting for different reasons. Indeed, it turns out that many, seemingly unrelated, well-defined four-dimensional theories and regularization techniques precisely admit the same spherically symmetric and FLRW classes of solution, as well as first order perturbations, as this ``minisuperspace regularization".
    
Indeed, both the Mann-Ross conformal regularization \cite{ConfGB2,ConfGB1,ConGB12} and the Kaluza-Klein reduction of the theory \cite{HornGB1,HornGB2} yield a shift-invariant Horndeski theory admitting a branch of solutions that agree with the ``minisuperspace regularization". Furthermore, the same scalar-tensor theory has been found from the study of renormalization group flows in \cite{RenGroupAnomCCRGB}. Similarly, the same regularized sector of Gauss-Bonnet gravity has been obtained in the context of Ho\v{r}ava-Lifshitz gravity in \cite{Horava} and from a direct regularization of the former theory in \cite{3Dcov1,3Dcov2,3Dcov3}, resulting in a theory which belongs to the class of Non-minimally modified gravity, i.e. 3D-covariant theories propagating only the two degrees of freedom of the graviton, see \cite{MMG1,MMG2,MMG3,MMG4,MMG5}. 
Finally, the regularized Gauss-Bonnet black hole also appeared as a solution of the semi-classical Einstein equations with conformal anomaly in \cite{conformal}.

From these examples, it seems quite natural to investigate first if the ``minisuperspace regularization" of the entire Lovelock-Lanczos series is also well-defined and if it yields to the resolution of specific issues of GR, like its singularities. If so, the next step would be to find well-defined four-dimensional theories admitting these previously found Lovelock sectors (as, for example, the shift-invariant Horndeski model discussed above). Possible candidates to achieve this have been found in \cite{DALLG}, where many background independent regularizations of Lovelock-Lanczos gravity have been found.

In this paper, we are interested in the first point so we extend the previous results of the ``minisuperspace regularization" for Lovelock-Lanczos terms of arbitrary curvature order, focusing on the previously mentioned sectors for which the limit $d\to 4$ has been shown to be well-defined in the case of Gauss-Bonnet gravity. We study the first order perturbations around (A)dS$_4$ in Sec. \ref{sec:AdeSitter}, the cosmological solutions with arbitrarily curved spatial sections in Sec. \ref{sec:FLRW}, and the spherically symmetric solutions in Sec. \ref{sec:SSS}. Note that the generalization to higher order Lovelock-Lanczos terms and the associated black hole solutions and thermodynamics has also been investigated very recently in \cite{higherorderLovelockblackholes}.

In the second part we focus on the truncated regularized Lovelock-Lanczos theory at cubic order in the series. We find the solutions of the Einstein equations for a curved FLRW metric, in the presence of a generic barotropic fluid. Restricting to the flat FLRW case, by using a stability analysis, we find that de Sitter critical points are always stable attractors of the evolution. The behaviour is similar to what happen with GR. The main difference lies on the fact that the de Sitter points can be generated also with vanishing cosmological constant, and they live in different disconnected sectors of the space of solutions. Regular dS vacuum solutions are explicitly found. We find that, as in GR, bouncing solutions are allowed only in closed spacetimes.

Finally, in Sec. \ref{sec:BHsol}, a one-parameter family of black hole solutions is found in the cubic case. They constitute a very natural generalization of the quadratic solution, and, although they possess a curvature singularity, the divergence of curvature invariants is milder than in the standard case. This class of solutions is then generalized to arbitrary high order truncation, allowing us to deduce that the central singularity of these black holes disappear in the non-perturbative limit.

Throughout the whole paper we consider $c = 1$ and $16 \pi G = 1$.

\section{Regularized Lovelock gravity}\label{sec:lovelock}

In this section we review some general results on the Lovelock-Lanczos theory, and introduce the regularization. We consider the following action in $d$ dimensions
\begin{equation}
    S = \int d^dx \,\sqrt{-g} \, \left[\sum_{p=0}^t \alpha_p \, \mathcal{L}_p\left(R,R_{\mu \nu},R_{\mu \nu \rho \sigma}\right)\right]\, + S_m\,,\label{eq:action_g}
\end{equation}
where the first part is the Lovelock--Lanczos action in $d$ dimensions, and the second is the term containing the matter contribution. The parameter $t$ is called the order of the Lovelock gravity, $\alpha_p$ are coupling constants and $\mathcal{L}_p$ are functions of the curvature scalar and tensors, which are defined as
\begin{equation}
\mathcal{L}_p = \frac{1}{2^p} \delta^{\mu_1 \nu_1 \dots \mu_p \nu_p}_{\sigma_1 \rho_1 \dots \sigma_p \rho_p} \prod  _{r=1}^p R^{\phantom{\mu_r \nu_r} \sigma_r \rho_r}_{\mu_r \nu_r}\,,
\end{equation}
where $\delta^{\mu_1 \nu_1 \dots \mu_p \nu_p}_{\sigma_1 \rho_1 \dots \sigma_p \rho_p}$ is the generalized Kronecker delta. Note that the coupling constants $\alpha_p$ have mass dimension $[M]^{2(p-1)}$.
The explicit definitions for $\mathcal{L}_p$ with $p=1,2,3$  are
\begin{align}
\mathcal{L}_0 &= 1 \,,\label{eq:L_0}\\
\mathcal{L}_1 &= R \,,\label{eq:L_1}\\
\mathcal{L}_2 &= R^2 - 4 R^{\mu \nu} R_{\mu \nu} + R^{\mu \nu \rho \sigma} R_{\mu \nu \rho \sigma} \equiv \mathcal{G}\,, \label{eq:L_2}\\
\mathcal{L}_3 &= R^3   -12RR_{\mu \nu } R^{\mu \nu } + 16R_{\mu \nu }R^{\mu }_{\phantom{\mu } \rho }R^{\nu \rho }+ 24 R_{\mu \nu }R_{\rho \sigma }R^{\mu \rho \nu \sigma }+ 3RR_{\mu \nu \rho \sigma } R^{\mu \nu \rho \sigma } \nonumber \\
&-24R_{\mu \nu }R^\mu _{\phantom{\mu } \rho \sigma \kappa } R^{\nu \rho \sigma \kappa  }+ 4 R_{\mu \nu \rho \sigma }R^{\mu \nu \eta \zeta } R^{\rho \sigma }_{\phantom{\rho \sigma } \eta \zeta }-8R_{\mu \rho \nu \sigma } R^{\mu  \phantom{\eta } \nu }_{\phantom{\mu } \eta 
\phantom{\nu } \zeta } R^{\rho  \eta  \sigma  \zeta }\,. \label{eq:L_3}
\end{align}
In particular, note that Eq. \eqref{eq:L_2} corresponds to the Gauss-Bonnet term $\mathcal{G}$. The case $t=1$ corresponds to the Einstein-Hilbert action with a cosmological constant $\Lambda_0$ provided we set $\alpha_0=-2\Lambda_0$ and $\alpha_1 = 1$. 

The field equations of the theory can be computed varying the action in Eq. \eqref{eq:action_g} with respect to the metric $g_{\mu \nu}$. The result is
\begin{equation}
\mathcal{G}_{\mu \nu} = \sum_{p=0}^t \alpha_p \, \mathcal{G}^{(p)}_{\mu\nu} = \frac{1}{2}\, T_{\mu\nu},
\label{eq:eom_t}
\end{equation}
where $T_{\mu\nu}$ is the stress-energy tensor, coming from the variation of the matter part $S_m$ of the action in Eq. \eqref{eq:action_g}, and 
\begin{equation}
\label{eq:Lovelocktensor}
\mathcal{G}_\beta^{(p) \alpha}\big[g_{\mu\nu} \big]  = \frac{g^{\alpha\gamma}}{\sqrt{-g}}\frac{ \delta\left[\sqrt{-g} \mathcal{L}_p \right]}{\delta g^{\gamma\beta}} =  -\frac{1}{2^{p+1}} \delta^{\alpha \mu_1 \nu_1 \dots \mu_p \nu_p}_{\beta \sigma_1 \rho_1 \dots \sigma_p \rho_p}\prod  _{r=1}^p R^{\phantom{\mu_r \nu_r} \sigma_r \rho_r}_{\mu_r \nu_r} \, .
\end{equation} 
By construction, Eq. \eqref{eq:Lovelocktensor} identically vanishes for $d \leq 2p$, because of the totally antisymmetric Kronecker delta in its definition.

In the following, we focus on some specific parametrizations of the metric field: the (A)dS metric and the first order perturbations around it, the FLRW sector as well as static spherically symmetric gravitational fields. By first making a $d$-dependent rescaling of the coupling constants $\alpha_p$ to implement the  regularization discussed above, and fixing the metric field as one of the previous ansatz, we find non-vanishing contributions to the field equations from generic Lovelock terms when the dimension of the manifold is fixed at the end of the procedure.

\subsection{(Anti) de Sitter solutions and perturbations}\label{sec:AdeSitter}

In this section we will discuss the (anti)-de Sitter solutions of Eq.\ \eqref{eq:eom_t} and the corresponding first order perturbation equations. We  take into account the possibility that the Lovelock terms produce an effective cosmological constants $\Lambda$ which differs from  $\Lambda_0$, which we will call bare cosmological constant \eqref{eq:action_g}.

For an (anti)-de Sitter spacetime, denoted with a metric $\bar{g}$, the components of the Riemann tensor reduce to the maximally symmetric form
\begin{equation}
R_{\mu\nu\sigma\rho}  = \Lambda \left(  \bar{g}_{\mu\sigma}  \bar{g}_{\rho \nu}  -   \bar{g}_{\mu \rho}  \bar{g}_{\sigma \nu} \right) \,.
\end{equation}{}
The field equations of each Lovelock scalar simplifies to 
\begin{equation}
\mathcal{G}_{\mu}^{(p)\nu} \left[ g=  \bar{g}  \right] = -\frac{1}{2} \frac{(d-1)!}{(d-2p-1)!} \Lambda^{p} \delta_\mu^\nu \,.
\label{eq:eom_AdSp}
\end{equation}{}
Therefore the (A)dS vacua can be obtained by solving the polynomial equation
\begin{equation}
 \Lambda_0 -\frac{(d-1)(d-2)}{2}  
 \Lambda -\frac{1}{2}  \sum_{p=2}^t \alpha_p \, \frac{(d-1)!}{(d-2p-1)!} \Lambda^{p}  =0  \, .
\end{equation}{}
As expected, these contributions vanish for $d \leq 2p$. However, if we apply the regularization, i.e. if one redefines the coupling constants $\alpha_p$ as  
\begin{equation}
\alpha_p = \frac{(d-2p-1)!}{(d-3)!} \tilde{\alpha}_p  \,,
\end{equation}
the field equations are no longer vanishing for $d \leq 2p$, because the poles in $\alpha_p$ cancel precisely the zeros of \eqref{eq:eom_AdSp}. Therefore, by considering the limit $d\rightarrow 4$ only at the end of the computation, it is possible for the Gauss-Bonnet and the $\mathcal{L}_3$ terms to introduce non-trivial contributions to the equations of motion. The same is true for any $p\geq 2$. For instance, setting $\Lambda_0=(d-1)(d-2) \tilde{\Lambda}/2$, the equations of motion for $t=3$ substantially simplify and read
\begin{equation}
\frac{(d-1)(d-2)}{2}\left( \tilde{\Lambda} - 
 \Lambda -  \tilde{\alpha}_2 \, \Lambda^2  -  \tilde{\alpha_3} \, \Lambda^3 \right)=0  \,, \label{eq:eom_t3_redef}
\end{equation}
where all the dimensional dependence of this field equation is contained in its pre-factor. Because of this property, from now on we consider $\tilde{\Lambda}$ as the bare cosmological constant, and set $\tilde{\alpha}_2 = \alpha$, $\tilde{\alpha}_3 = \beta$. Note that, in the GR limit ($\alpha=\beta=0$) we consistently find $\Lambda =\tilde{\Lambda}$.

From Eq. \eqref{eq:eom_t3_redef} we can also find the relation between the effective cosmological constant $\Lambda$ and the bare cosmological constant $\tilde{\Lambda}$. For example, in the case $\beta=0$, we obtain two possible values
\begin{equation}
 \Lambda_{\pm} = -\frac{1}{2 \alpha} \left( 1 \pm \sqrt{1 + 4 \alpha \tilde{\Lambda}} \right)
\end{equation}
which, in the case $\tilde{\Lambda}=0$ $(=\Lambda_0)$, gives two solutions: $\Lambda=0$ and $\Lambda = -1/\alpha$. Therefore, we find that an (anti)-de Sitter solution to the generalized equations of motion exists, even in the absence of a bare cosmological constant. \footnote{Note that the existence of (anti)-de Sitter solutions in modified gravity theories should not be taken for granted, see for example \cite{Casa19}.} In particular, if $\alpha<0$ and small, we have a de Sitter solution with a very large effective cosmological constant, which recovers GR at low curvature.

We now focus on the derivation of the linearized perturbation equations. The background solution is the (A)dS solution found above. We set
\begin{equation}
g_{\mu \nu} = \bar{g}_{\mu \nu} + h_{\mu \nu}\,,
\end{equation}
where $h_{\mu \nu}$ represents the first order perturbation to the (A)dS background $\bar{g}_{\mu \nu}$. The perturbed equations of motions read
\begin{equation}
\mathcal{G}_{\mu}^{(p)\nu} \left[g= \bar{g} + h \right] =  \frac{p}{2} \, \frac{(d-3)!}{(d-2p-1)!} \Lambda^{p-1} P_{\mu}^{\nu} \, ,
\end{equation}{}
where the tensor $P$ is defined by
\begin{align}
P^{\nu}_{\mu} =&\nabla_\mu \nabla_\rho h^{\nu \rho} + \nabla^\rho \nabla^\nu h_{\mu\rho} -\square h^\nu_\mu -\nabla^\nu\nabla_\mu h + \nonumber\\
&+\delta_\mu^\nu \left(\square h - \nabla_\sigma \nabla_\rho h^{\rho\sigma}\right) + (d-2) \Lambda \left(\delta_\mu^\nu h - h^\nu_\mu\right) \,.
\end{align}
where $h=h^{\mu}_{\mu}$.
Therefore, it is still possible to use the previous rescaling of the coupling constant $\tilde{\alpha}_p$, so that the first order field equations are not vanishing for $d \leq 2p$
\begin{equation}
\frac{1}{2} \left( 1 + \sum_{p=2}^t p \, \tilde{\alpha}_p \, \Lambda^{p-1} \right) P_{\mu}^{\nu}=0 \, .
\end{equation}
As the tensor $P$ is independent from the order $p$ of the Lovelock scalars,
this shows that the perturbation spectrum is indistinguishable from the one of GR, at least at the first order in the perturbations and provided that $1 + \sum_{p=2}^t p \,  \tilde{\alpha}_p \, \Lambda^{p-1} \neq 0$. This result is in accordance with the findings of \cite{Glavan:2019inb} for the case $t=2$.

On the other hand, if the coupling constants are such that $1 + \sum_{p=2}^t p \,  \tilde{\alpha}_p \, \Lambda^{p-1} = 0$, the theory does not admit a linear dynamics around the (A)dS vacuum $\Lambda$, meaning that it is strongly coupled. Such couplings define critical points in Lovelock theory space, and are related to the degeneracy of the corresponding vacua, see \cite{DALLG} for the full classification of these degeneracies around (A)dS and \cite{TekinCrit,LUV} for the conserved charges associated with such theories.

\subsection{Friedmann-Lema\^{i}tre-Robertson-Walker solutions}\label{sec:FLRW}

In this section we focus on the equations of motion of Lovelock gravity for a curved FLRW spacetime. We consider the metric
\begin{equation}\label{RWmetric}
ds^2 = - dt^2 + a(t)^2 g^{(n+1)}_{ij} dx^i dx^j = - dt^2 + a(t)^2 \left(\frac{dr^2}{1-kr^2} + r^2 d\Omega_n^2\right)\,,
\end{equation}
where $d\Omega_n$ is the line element of the $n=d-2$ hyper-sphere; $a(t)$ is the scale factor; $k$ is the space curvature ($k=0$ is the flat space, while $k>0,k<0$ describes respectively a closed and an open geometry for the spatial sector).
With this metric we obtain 
\begin{equation}
 \mathcal{G}_{0}^{(p)0}=-\frac{1}{2} \frac{(d-1)!}{(d-2p-1)!} J^{2p} \,\; , \;\;\;  \mathcal{G}_{0}^{(p)0}-  \mathcal{G}_{1}^{(p)1} = p  \, \frac{(d-2)!}{(d-2p-1)!} J^{2(p-1)} \left( Q^2 - J^2 \right)  \, ,
\end{equation} 
where we defined 
\begin{equation}
    J^2 = H^2 + \frac{k}{a^2}\,,\qquad Q^2=H^2+\dot{H}\,.
\end{equation}
Here $H(t) = \dot{a}(t)/a(t)$ is the Hubble parameter and the dot stands for the derivative with respect to the cosmological time coordinate $t$. These terms have the special property that, if evaluated on de Sitter manifolds, they are invariant under change of $k$. Indeed, the de Sitter solution with metric \eqref{RWmetric} yields $a(t)=\exp(H_0t)$ for $k=0$, $a(t)=\cosh(H_0t)$ with $k=H_0^2$, and $a(t)=\sinh(H_0t)$ with $k=-H_0^2$. In all cases the Ricci scalar is $R=12H_0^2$ and, more importantly, $J^2=Q^2=H_0^2$ for all $k$ \cite{DiCriscienzo:2009hd,Acquaviva:2011vq}.

Once again, we see that the field equations vanish for $d \leq 2p$. However, in terms of the rescaled constants $\tilde{\alpha}_p$, the $d$-dimensional Friedmann equations become 
\begin{align}
& \mathcal{G}_{0}^{0}=\Lambda_0 - \frac{(d-1)(d-2) }{2} \sum_{p=1}^t \tilde{\alpha}_p J^{2p} = - \frac{\rho}{2} \,,\label{eq:fried1_t}\\
&              
  \mathcal{G}_{0}^{0}-  \mathcal{G}_{1}^{1} =  -(d-2) \left(J^2-Q^2\right) \sum_{p=1}^t  p \, \tilde{\alpha}_p \, J^{2(p-1)}  =  -\dfrac{\left(\rho + P\right)}{2}\,,\label{eq:fried2_t}
\end{align}
of which we will consider the limit $d\rightarrow4$.  In the Friedmann equations above we introduce a minimally coupled perfect fluid with energy density $\rho$ and pressure $P$, which obeys the usual conservation equation in $d$ dimensions
\begin{equation}
    \dot{\rho} +  (d-1) H (\rho + P) = 0 \label{eq:cons}\,.
\end{equation}

\subsection{Spherically symmetric solutions}\label{sec:SSS}

Finally, we consider the equations of motion for a spherically symmetric spacetime. We consider the static metric
\begin{equation}
    ds^2 = - A(r) B(r)^2 dt^2 + \frac{1}{A(r)} dr^2 + r^2 d\Omega_{d-2}^2\,,  \label{eq:metricsss}
\end{equation}
where $\Omega_{d-2}$ is the metric of the $d-2$ hyper-sphere. This hyper-sphere can be spherical, hyperbolic or planar for respectively $k=1,-1,0$, and is denoted by a metric $\sigma_{ij}$, with indices $i,j=2, ..., d-1$, and a determinant $\sigma$. The contribution of each Lovelock scalar to the field equations is given by 
\begin{align}
& \sqrt{-g} \,\mathcal{G}^{(p)0}_{0} =-\frac{\sqrt{\sigma}}{2}  \frac{(d-2)!}{(d-2p-1)!} B \left( r^{d-2p-1}  \left( k-A \right)^{p}   \right)'  \,,\label{eq:sss1_t}\\
&              
 \sqrt{-g} \,\left(\mathcal{G}^{(p)0}_{0}- \mathcal{G}^{(p)1}_{1} \right) =- \sqrt{\sigma} p \frac{(d-2)!}{(d-2p-1)!} r^{d-2p-1} \left( k-A \right)^{p-1} A B' \,,\label{eq:sss2_t}
\end{align}
where the prime denotes the derivative with respect to the radial coordinate $r$. Therefore, using the rescaled coupling constants $\tilde{\alpha}_p$, the second equation yields 
\begin{equation}
 \sqrt{-g}\left(\mathcal{G}^{0}_{0}- \mathcal{G}^{1}_{1} \right) =- \sqrt{-\sigma} (d-3) A  \left[ \sum_{p=0}^t p \, \tilde{\alpha}_p r^{d-2p-1} \left(k-A\right)^{p-1} \right] B' =0 \, . \label{eq:sss1}
\end{equation}
If the term in square brackets vanishes we find  the special degenerate vacuum solution discussed  in \cite{Maeda}. If it does not vanish, we find  $B=1$, so that the equation \eqref{eq:sss1_t} can be integrated to give
\begin{equation}
-2\int_\mathcal{M} d^d x \sqrt{-g} \,\mathcal{G}^{0}_{0} =  (d-2) V \sum_{p=0}^t \tilde{\alpha}_p r^{d-2p-1}\left(k-A\right)^p =  M_{\text{GMS}}  \, ,  \label{eq:sss2}
\end{equation}
where $V$ is the volume of $\Omega_{d-2}$ and $M_{\text{GMS}}$ is the generalized Misner-Sharp quasi-local mass. Therefore, Lovelock black holes can be found in four dimensions via the procedure of rescaling the coupling constants, evaluating the field equations and then setting $d=4$. 

Finally, we compute the Wald entropy of these solutions in terms of $\tilde{\alpha}_p$, as there is in this case a departure from higher dimensional Lovelock gravity. first, the Wald entropy is defined as 
\begin{eqnarray}
 S_{W}=-2\pi \int_{\Omega_{d-2}} \sqrt{\sigma} \, \left( \frac{\partial \mathcal{L}}{\partial R_{\mu\nu\alpha\beta}} \right) \epsilon_{\mu\nu} \epsilon_{\alpha\beta} d^2 x \,,
\end{eqnarray}
with $\epsilon_{\mu\nu}$ the binormal to the horizon $\Omega_{d-2}$, which is defined by $\epsilon_{\mu\nu} \epsilon^{\mu\nu} =-2$, $\epsilon_{01}=-\epsilon_{10}=1$; the other components are vanishing. We recall that, in this expression, the derivative of the Lagrangian with respect to the Riemann tensor  is evaluated on-shell. Finally, $\sigma$ is the determinant of the induced metric on the horizon. 

For $B(r)=1$ and $A\left(r\right)$ solution of \eqref{eq:sss2} with $M_{\text{GMS}} \geq 0$,
\begin{eqnarray}
S_W=2 \pi V  \sum_{p=0}^t p \,  \tilde{\alpha}_p \, \frac{d-2}{d-2p} \, k^{p-1} r_H^{d-2p}\label{eq:wald_entropy_2}
\end{eqnarray}
where $r_H \left(M_{\text{GMS}} \right)$ is the horizon radius. 

Contrary to the previous cases, there is a remaining pole in the expression \eqref{eq:wald_entropy_2}, namely at the critical order $p=d/2$, that corresponds to a regularization of the Gauss-Bonnet scalar in four dimensions. If we consider the limit, we have 
\begin{equation}
S_W \left( p \to d/2 \right) =2\pi(d-2) V \, l^{d-2}  \left\lbrace \frac{d  }{2(d-2p)} +  \left[ -\frac{1}{2} + \log\left( \frac{r_H}{l} \right)^{d/2}  \right]    \right\rbrace + O\left(d-2p\right)
\end{equation}
where $\tilde{\alpha}_{p}=l^{2(p-1)}$ and $l$ is a length scale associated with this order of correction. In order to get a finite result we must therefore subtract the constant divergent term. But this does not seem a problem because, being a state function, the action is defined up to a constant. Then, the only non-constant term in the limit $p \to d/2$ is a logarithmic correction, which is in agreement with Eq. ($25$) of \cite{Cognola:2013fva} for $d=4$.

\section{Cubic and higher order Lovelock gravity in four dimensions}

In this section we mainly focus on the four dimensional cubic Lovelock gravity. In this case the Lovelock action \eqref{eq:action_g} is truncated at $t=3$, becoming
\begin{equation}
\label{eq:action}
S = \int_\mathcal{M} d^d x \sqrt{-g} \left[ -2 \Lambda_0 + R + \frac{\alpha} {(d-3)(d-4)}\,\mathcal{G} + \frac{\beta}{(d-3)(d-4)(d-5)(d-6)} \,\mathcal{L}_3 \right]\,.
\end{equation}
As the previous results show, it is possible to evaluate the field equations for a specific ansatz and then set $d=4$. At the end of this section, we will also investigate some arbitrary order black holes and assess their singularity behaviour.

\subsection{Dynamical analysis in the flat case}\label{sec:acc}

We found in Sec. \ref{sec:AdeSitter} that de Sitter solutions are allowed. In this section we want to assess their stability in a flat FLRW. To this end, we apply the standard dynamical system analysis to the Friedmann equations \eqref{eq:fried1_t}, \eqref{eq:fried2_t} in the limit $d\rightarrow 4$, obtaining
\begin{align}
J^2 \left(1 + \alpha J^2 +  \beta J^4\right) = \frac{1}{6}\, \left(2 \Lambda_0 + \rho\right)\,,\label{eq:friedm1}\\
\left(Q^2-J^2\right) \left(1+2 \alpha J^2 + 3 \beta J^4 \right)= -\frac{1}{4} \left(\rho + P\right)\,.\label{eq:friedm2}
\end{align}
For simplicity, from now on we assume that the fluid is barotropic such that  $P=\omega\rho$, with $\omega$ a constant equation of state parameter. Moreover, we focus on the flat case, i.e. the case $k=0$, where $J^2=H^2$ and $Q^2-J^2 = \dot{H}$.

We now study the fixed points and their stability of the equations above. In the flat case, equation \eqref{eq:friedm1} reads
\begin{equation}
H^2= \frac{1}{6}\, \left(2\Lambda_0 + \rho\right) - \alpha H^4 - \beta H^6\,.\label{eq:f1}
\end{equation}
We define 
\begin{equation}
x = H^2 \qquad \text{and} \qquad y = \frac{\rho}{6}\,,\label{eq:variables}
\end{equation}
so that equation \eqref{eq:f1} becomes
\begin{equation}
x = y + \frac{\Lambda_0}{3} -\alpha x^2 -\beta x^3\,.\label{eq:constr}
\end{equation}
Since this equation does not contain derivatives of the variables $x$ and $y$, it acts as a constraint. Similarly,  equation \eqref{eq:friedm2} in the flat case becomes
\begin{equation}
\dot{H} = -\frac{3}{2}(1+\omega)\frac{y}{1+2\alpha x+3\beta x^2}\,.\label{eq:f2}
\end{equation}
We can rewrite the equations using the e-folding number $N\equiv \ln a$. From the definition \eqref{eq:variables} we obtain
\begin{equation}
\frac{d}{dN} \ln(x) = \frac2H\, \frac{d H}{dN} = 2\, \frac{\dot{H}}{H^2}\,,
\end{equation}
which can be written as a function of the dynamical variables using Eq. \eqref{eq:f2}, obtaining
\begin{equation}
\frac{d x}{dN} = -3(1+\omega)\frac{y}{1+2\alpha x+3 \beta x^2}\,. \label{eq:diff_x}
\end{equation}
The second dynamical equation for $y(N)$ can be found using the definition of $y$ in Eq. \eqref{eq:variables}, and the perfect fluid equation \eqref{eq:cons}. We obtain
\begin{equation}
\frac{dy}{dN} = -3(1+\omega)y\,.\label{eq:diff_y}
\end{equation}
Finally, the system of equations \eqref{eq:diff_x}--\eqref{eq:diff_y} must also satisfy the constraint \eqref{eq:constr}, which yields to the single differential equation
\begin{align}
\frac{d x}{dN} &= -3(1+\omega)\frac{x - \frac{\Lambda_0}{3} +\alpha x^2 +\beta x^3}{1+2 \alpha x+3 \beta x^2}\label{eq:sys1}
\end{align}
Note that the equation is well-defined provided the further condition $
1+2\alpha x+3\beta x^2\neq 0$ holds. In particular, in the case $\beta=0$, the system is ill-defined if $x=-1/2\alpha$.

The stationary points can be found by setting $dx/dN=0$ in the equation above. If $\omega\neq -1$, they are given by the equation
\begin{equation}
x+\alpha x^2+\beta x^3 = \frac{\Lambda_0}{3}\,.\label{eq:x_sol}
\end{equation}
In general, if we increase the number of terms of the Lovelock series beyond the third order, we may have more stationary points. Moreover, if  the bare cosmological constant $\Lambda_0$ is null, $x=x_0=0$ is a stationary point for any order. In such a case, if $y=0$ (that is no matter content), this point corresponds to a Minkowski metric.

In particular, in the case of $\beta=0$, the explicit two solutions of \eqref{eq:x_sol} are given by
\begin{equation}
x_{\pm} = -\frac{3 \pm \sqrt{9+12\alpha\Lambda_0}}{6 \alpha}\,\label{eq:sp_a1}
\end{equation}
from which, when $\Lambda_0 = 0$, we obtain the points $x_-=0$, which corresponds the the Minkowski solution mentioned above, and $x_+=-1/\alpha$, that corresponds to the de Sitter solution provided $\alpha<0$. 

\begin{center}
\begin{figure}[ht]
\includegraphics[scale=0.2]{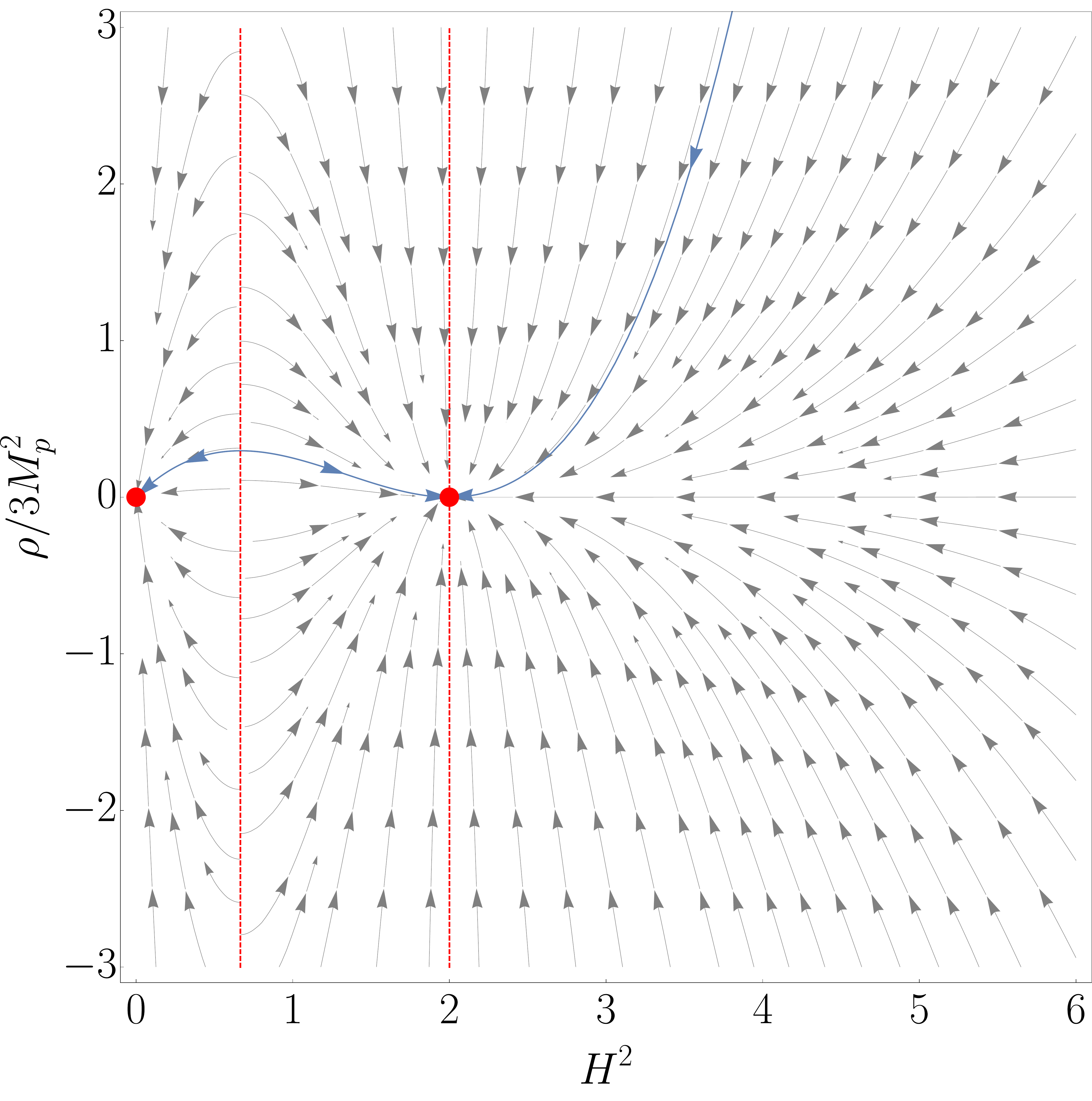}
\caption{Stream plot for the third-order case, with $\alpha=-0.5$, $\beta=1/4$, $\omega=0$ and $\Lambda_0=0$.  Two fixed points now coincide on a repellor line. Also in this case, the physical trajectory (blue line) is a  viable solution.}\label{fig:dyn3}
\end{figure}
\end{center}

The next step is to assess the stability of the stationary points. We first consider the $t=2$ ($\beta=0$) case and then generalize to higher orders below. We use the Hartmann-Grobman theorem and study the linearised differential equations near a stationary point $x_0$. Thus, we linearise the right-hand-side of equation \eqref{eq:sys1} and change variable to $\tilde{x} = x - x_0 = x - x_{\pm}$. The result is
\begin{equation}
\frac{dx}{dN} =-3(1+\omega) \frac{\sqrt{9+12\alpha\Lambda_0} \mp 3 \alpha \tilde{x}}{\sqrt{9+12\alpha\Lambda_0} \mp 6 \alpha \tilde{x}} = -3(1+\omega) \tilde{x} + \text{o}[ \tilde{x}^2]\,.\label{eq:dyn_lin}
\end{equation}
Since $-3(1+\omega)$ is always negative for $\omega>-1$  the stationary points $x_\pm$ are both stable attractors. The result can be easily generalized to higher order of the Lovelock series. In fact, the new terms introduce higher powers of $\tilde{x}$, leaving the linear one unchanged. Thus, we find that at order $t$ of the Lovelock series with $\Lambda_0=0$, the system has always $t$ fixed points (which might also coincide, see e.g. see Fig. \ref{fig:dyn3}), which are stable attractors.

\begin{center}
\begin{figure}[!htp]
\includegraphics[scale=0.2]{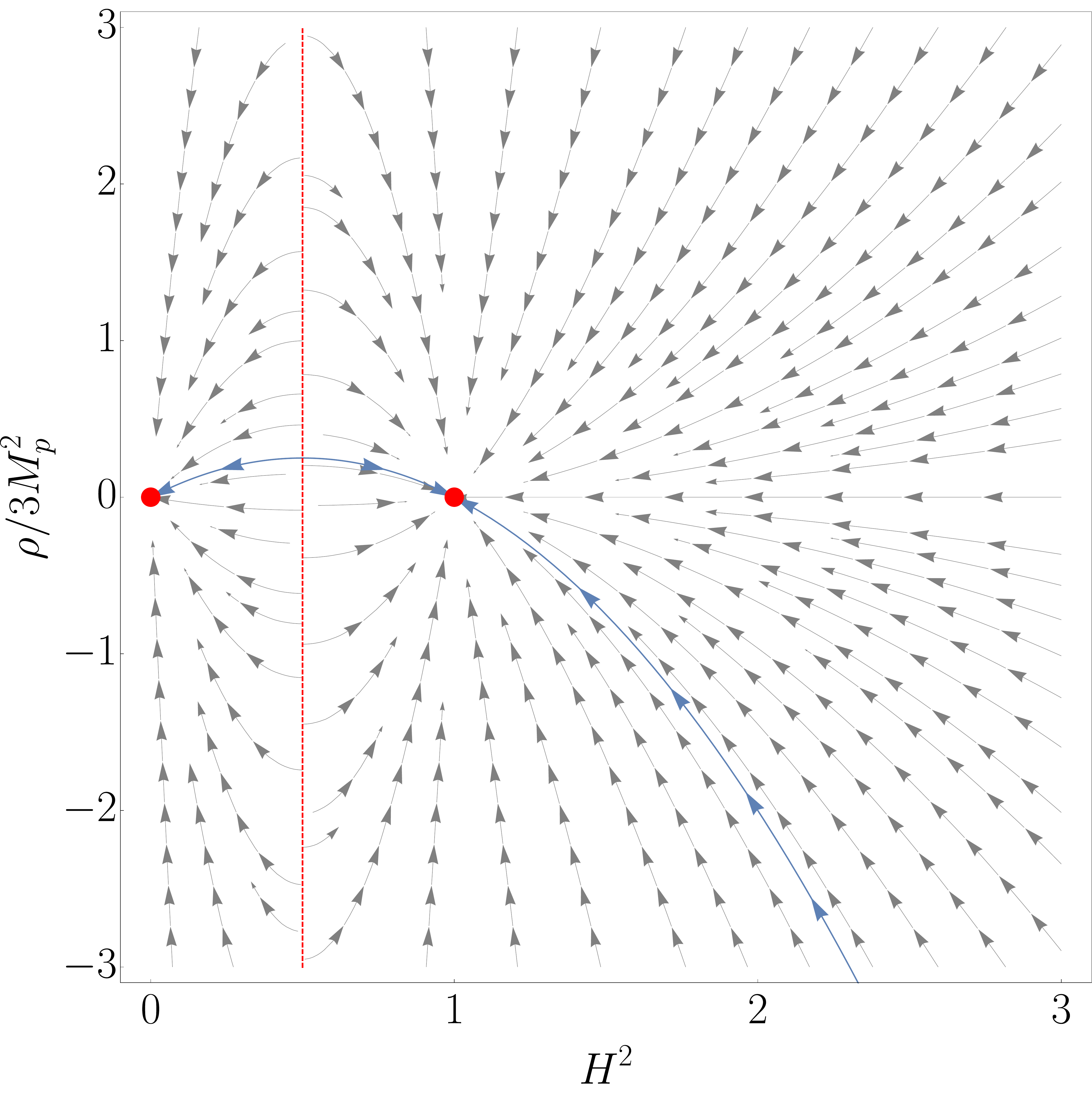}
\caption{Stream plot for $\beta=0$, with $\alpha=-0.5$, $\omega=0$ and $\Lambda_0=0$. $N$ increases in the directions of the arrows, the red points are the stationary points and the red line corresponds to $x=H^2=-1/2\alpha$. The blue curve is the physical solution of the differential equation \eqref{eq:sys1} for any initial condition. We can note that $H$ increases only in the region where $\rho$ decreases}\label{fig:dyn1}
\end{figure}
\end{center}

\begin{center}
\begin{figure}[ht]
\hspace*{-0.5cm}\includegraphics[scale=0.2]{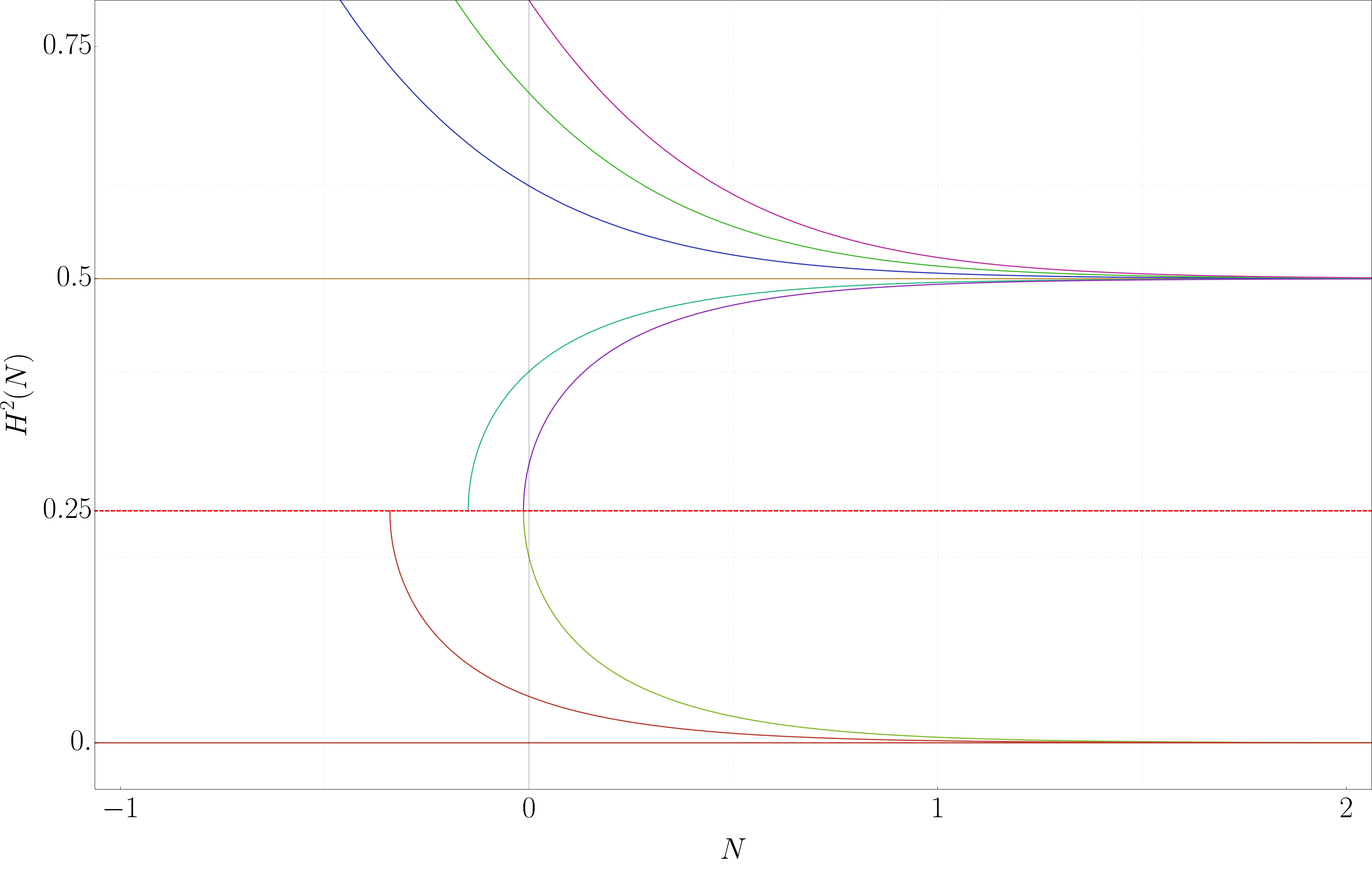}
\caption{Plot of the solutions of \eqref{eq:diff_x} as a function of the number of e-foldings $N=\text{ln}(a)$. We choose $\alpha=-0.5$, $\beta=0$, $\omega=0$ and $\Lambda_0=0$. The different solutions have different initial conditions for $H^2(N=0)$ (whose value can be read directly from the plot). The dashed line represents the unacceptable values for $H^2$ given by the equation $H^2=-1/2\alpha$. Note that the solutions with initial conditions on the attractor remain constant, while the other solutions evolve towards the attractor.}\label{fig:x_sol}
\end{figure}
\end{center}

We compute some numerical results for the quadratic Lovelock theory case ($\beta=0$). The stream plot for $\omega= \Lambda_0=0$ is reported in Fig. \ref{fig:dyn1} . 
The trajectories flow towards the two stable stationary points. The red line corresponds to $x=-1/2\alpha$, where the system is ill-defined (see Eq. \eqref{eq:sys1} and related discussion). We see that the trajectories are always repelled by this line and so there are two disconnected solution regions. Indeed if the initial value of $x$ is less than $-1/2 \alpha$, the solution will eventually end in the Minkowski stable point in finite time. On the contrary, if it originates from $x>-1/2 \alpha$, it eventually ends on the de Sitter stable point.
In Fig. \ref{fig:x_sol} we show some solutions of the differential equation \eqref{eq:diff_x} for different initial conditions $x(N=0)$.

\begin{center}
\begin{figure}[ht]
\includegraphics[scale=0.2]{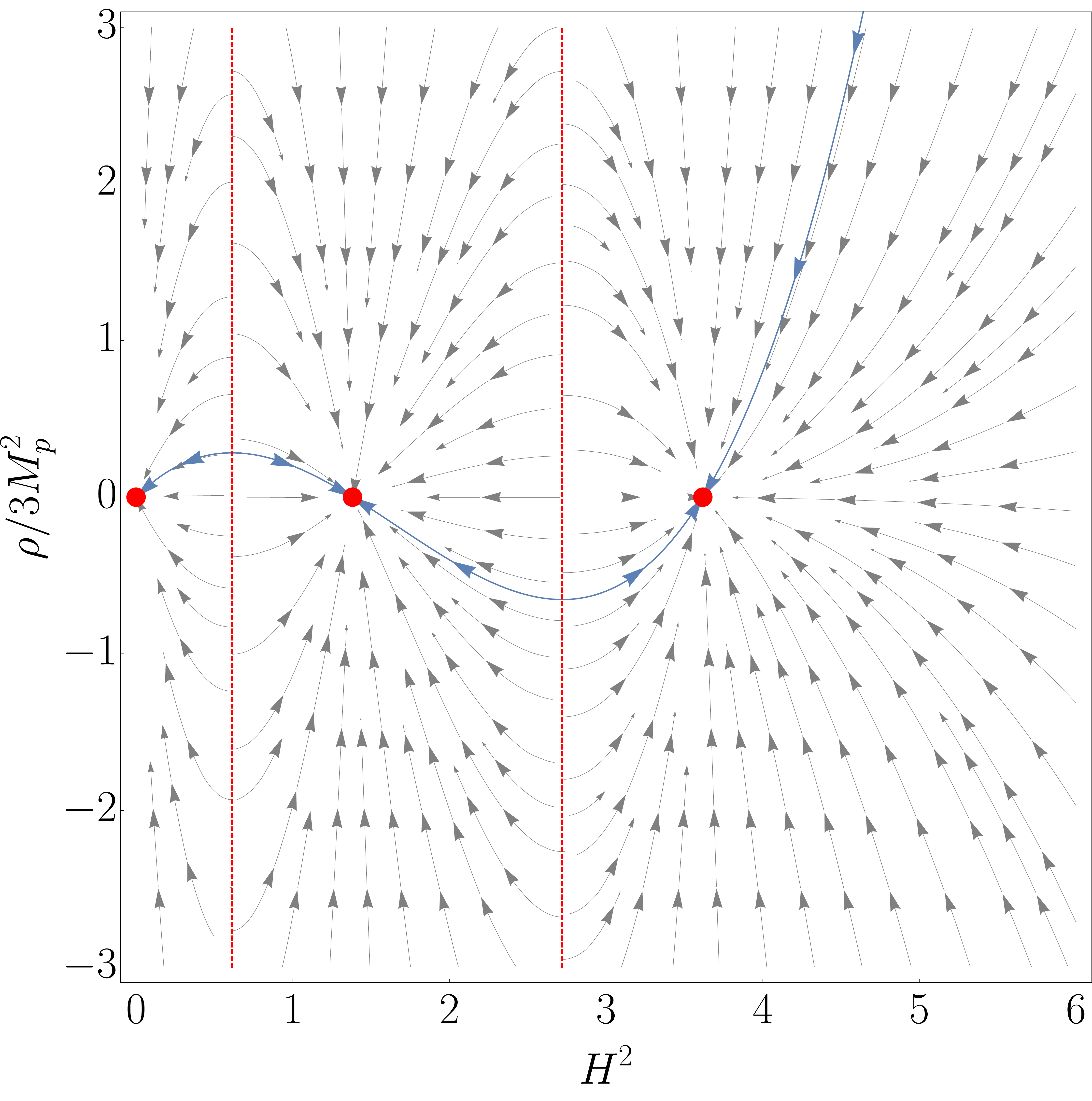}
\caption{Stream plot for the third-order case, with $\alpha=-0.5$, $\beta=1/5$, $\omega=0$ and $\Lambda_0=0$ ($M_p=1$), with matter perfect fluid ($\omega=0$). In this case, the right-region the physical trajectory (blue line) represents a viable solution with both $H$ and $\rho$ decreasing in time.}\label{fig:dyn2}
\end{figure}
\end{center}

If higher-order Lovelock terms are taken in account we find a different behaviour. For instance, if we consider the case $\beta\neq 0$, the procedure to find and study the stationary points is the same. As mentioned above, all the stationary points are stable in the future. From the algebraic equations of the stationary points, the analogous of Eq. \eqref{eq:sp_a1} in the case $\beta\neq 0$, when $\Lambda_0=0$, we see that we should also impose the condition $\beta \leq  \alpha^2 /4$ in order to have real stationary points.
The unacceptable values of $x$ (see Eq. \eqref{eq:sys1} and the related discussion) in this case are
\begin{equation}
x_{1,2} =- \frac{\alpha \pm \sqrt{\alpha^2 - 3 \beta}}{3 \beta}\,.\label{eq:unacc_3}
\end{equation}
When the condition $\beta \leq  \alpha^2 /4$ is satisfied,  $x_{1,2}$ are real numbers. If, in addition to this condition, we also consider $\alpha<0$ and $\beta>0$, then $x_{1,2}>0$. Although the explicit equations for the stationary points are straightforward to obtain, in order to avoid reporting cumbersome formulas, we limit ourselves to show the plot of the phase space in Fig. \ref{fig:dyn2}.
The case $\beta =  \alpha^2 /4$ deserves a particular attention. We have two coincident stationary points at $x=2/a_1$ (besides the usual one at $x=0$) while the unacceptable values are at $x=-2/(3\alpha)$ and $x=-2/\alpha$. Note that the latter coincides with the standard stationary point. In Figure \ref{fig:dyn3} we show the phase space in this particular case. We see that the coincident fixed points are still future attractors and that the solution trajectory is still physically viable since it indicates a decreasing $H$ together with a decreasing energy density $\rho$. 

From a cosmological point of view then, truncations of the Lovelock series may predict the current expansion of the Universe with a very large effective cosmological constant even if the bare cosmological constant is set to zero. In fact, for instance in the case $\beta=0$, the effective cosmological constant is given by $x_- = -1/\alpha$, which can be arbitrarily large if $\alpha$ is small as expected from considering the Gauss-Bonnet term as a small correction to GR. It is unclear if a particular combination of the coupling constants in the case of $t \gg 3$ might lead to a small effective cosmological constant.

\subsection{Vacuum solutions}\label{sec:cosmBounce}

In this section we study the Friedmann equations \eqref{eq:fried1_t} and \eqref{eq:fried2_t} in the absence of any type of matter. We find that bouncing solutions exist in the case of positive  curvature, even if the bare cosmological constant $\Lambda_0$ is set to zero.

We have two classes of solutions. The first is
\begin{equation}
Q^2 = J^2 \qquad \text{and} \qquad 1 + \alpha J^2 + \beta J^4 =0\,,\label{eq:cond1}
\end{equation}
from which we get
\begin{equation}
J ^ 2 = - \frac{\alpha}{2 \beta} \left(1 \pm \sqrt{1 - \frac{ 4 \beta}{\alpha}}\right) = Q^2\,,
\end{equation}
where we assume $\beta\leq \dfrac{\alpha}{4}$. Therefore the first equation in \eqref{eq:cond1} can be written as   $\ddot{a}(t) = J^2 a(t)$, with $J^2=$ const. By choosing an appropriate initial condition we find a bouncing solution
\begin{align}
    a(t) &= a_0 \, \text{cosh} \left( \left|J\right| t \right)\,,\\
    H(t) &= J \, \text{tanh} \left(\left|J\right| t \right)\,,
\end{align}
where $a_0=\sqrt{k}/J$ is the integration constant fixed by the condition $H^2 +k/a^2 = J^2$. The condition $k>0$ follows and shows that a cosmological bounce is possible if the spatial curvature is positive. In fact, this is the usual bounce of the de Sitter patch with $k>0$.
    
The second class of solutions is
    \begin{equation}
        1 + \alpha J^2 +  J^4 = 0 \qquad \text{and} \qquad 1+2 \alpha J^2 + 3 \beta J^4 = 0\,,
    \end{equation}
    with $J^2 \neq Q^2$. We find (for $\beta>0$)
    \begin{equation}
        J = \pm \beta^{-1/4} \qquad \text{with} \qquad \alpha = - 2\sqrt{\beta}<0\,.
    \end{equation}
    This gives a differential equation for the scale factor
    \begin{equation}
        \frac{\dot{a}^2}{a^2} = J^2 - \frac{k}{a^2}\,,
    \end{equation}
    which has a regular bouncing solution
    \begin{equation}
        a(t) = \frac{\sqrt{k}}{J} \text{cosh}\left(\left|J\right| t\right)\,,
    \end{equation}
    for $k>0$ as for the previous case, but for a different value of  $J^2$.

In conclusion this model admits cosmological bouncing solutions, which corresponds to the the Sitter solution with positive spatial curvature, even when the bare cosmological constant vanishes.
As for the fixed points studied in the previous sections, cosmological bounces occur only for certain ranges of the parameters $\alpha$ and $\beta$ for the truncated series. In the case of infinite terms of the Lovelock series, the bouncing solutions simply corresponds to the  de Sitter fixed point with $k>0$, if they exists (namely if the polynomial function of $J^2$ in Eq. \eqref{eq:fried1_t} has at least one positive zero).

\subsection{Black hole solutions}\label{sec:BHsol}

In this section we start by considering black hole solutions in the case $t=3$. To this end, consider the line element Eq. \eqref{eq:metricsss}. Setting $d=4$, we can evaluate  Eq. \eqref{eq:sss1} and find
\begin{equation}
 \sqrt{-g}\left(\mathcal{G}^{0}_{0}- \mathcal{G}^{1}_{1} \right) =- \sqrt{-\sigma}  A   \left[  r + 2  \alpha \frac{k-A}{r} + 3 \beta  \frac{\left(k-A\right)^2}{r^3} \right] B'= 0 \,.
\end{equation}
Therefore, if term in the square brackets is not null, we can set $B=1$ without loss of generality. This reduces the equation for the generalized Misner--Sharp quasi-local mass, Eq. \eqref{eq:sss2}, to
\begin{align}
-2\int_\mathcal{M} d^4 x \sqrt{-g} \, \mathcal{G}^{0}_{0} &= 2V\left[ -\frac{\Lambda_0}{3} r^3 + r \left(k-A\right) + \alpha  \frac{\left(k-A\right)^2}{r} + \beta \frac{ \left(k-A\right)^3}{r^3} \right] \nonumber\\
 &=    M_{\text{GMS}}\,.
\end{align}
Considering a truncation of Lovelock theory at quadratic order ($\beta=0$), we obtain the black hole solution found in \cite{Tomozawa:2011gp,Cognola:2013fva}
\begin{equation} \label{eq:a2r}
A^{(2)}_\pm(r) = k - \frac{r^2}{2\alpha} \left[ -1 \pm  \sqrt{1+ 4 \alpha \left( \frac{2M}{r^3} +\frac{\Lambda_0}{3} \right)} \right] \,.
\end{equation}
where $M=M_{GMS}/4 V$.
Similarly, it is possible to find a generalization of this black hole considering the next curvature correction and setting $\beta = \alpha^2/3$ in order to have a single real vacuum,
\begin{equation}
A^{(3)}(r) = k - \frac{r^2}{\alpha} \left[ -1 +   \sqrt[3]{  1+ 3 \alpha\left( \frac{2M}{r^3}  + \frac{\Lambda_0}{3} \right) }\right]\,.\label{eq:A3}
\end{equation}
For spherical topology ($k=1$) and positive bare cosmological constant $\Lambda_0$, this solution describes a black hole with Schwarzschild-de Sitter behaviour at $r\to \infty$. 

\begin{center}
\begin{figure}[ht]
\hspace*{-1cm}\includegraphics[scale=0.20]{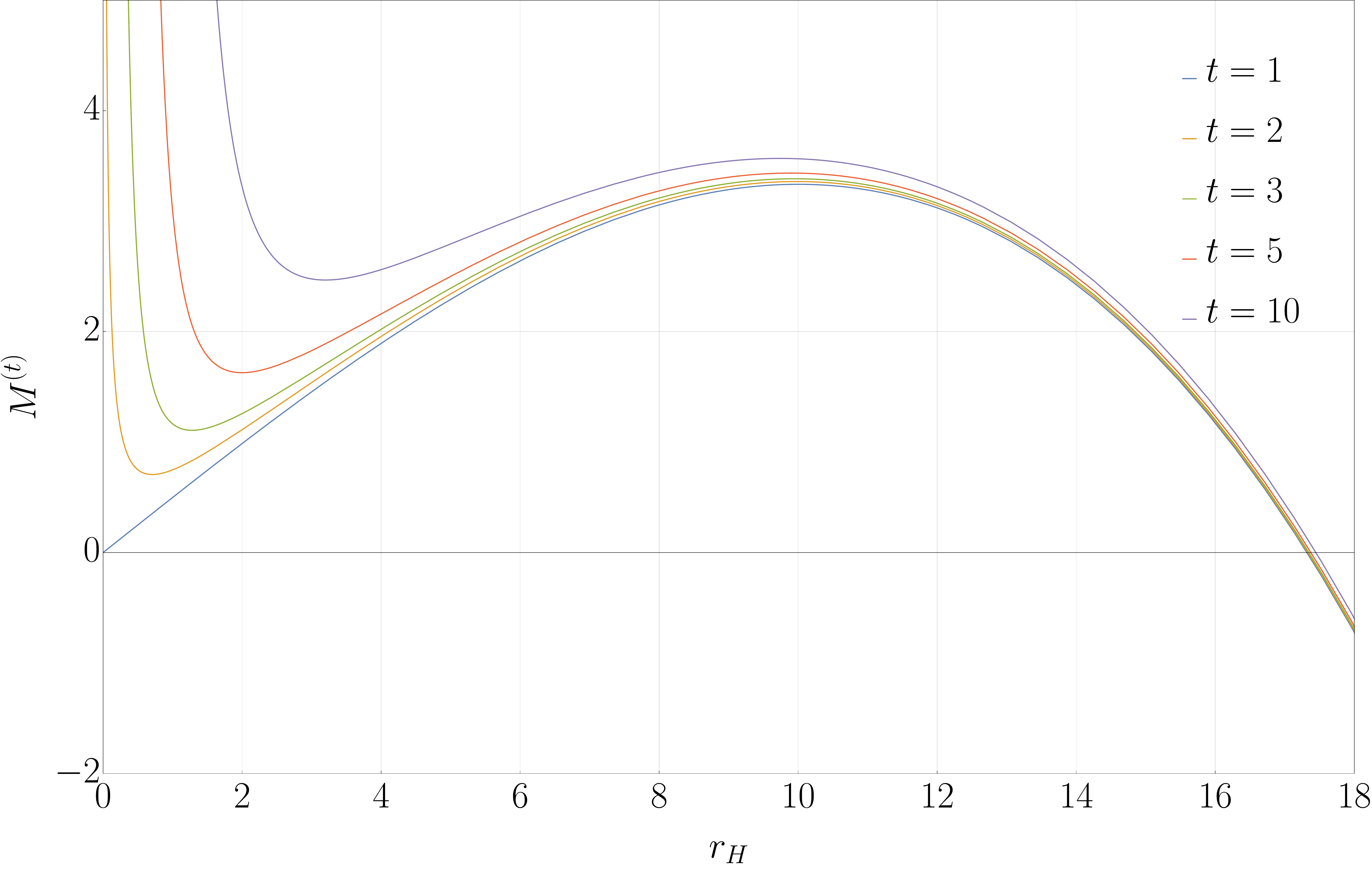}
\caption{Plot of the mass of the black hole, as a function of the horizon radius $r_H$, for different values of the Lovelock order parameter $t$. The equation for the mass is obtained inverting $A^{(t)}$ in Eq. \eqref{eq:At} evaluated at the horizon radius $r_H$. In this plot we set $\alpha=1$, $M=1$, $\Lambda_0=0.1$. \label{fig:Mt}}
\end{figure}
\end{center}

Looking at the mass function $M(r_H)$ (see the case $t=2$ in Fig. \ref{fig:Mt}), we see that, unless extremal, the black hole possesses three horizons for generic values of $\alpha$, $M$ and $\Lambda_0$: a cosmological horizon (due to $\Lambda_0 \neq 0$), as well as an event and an inner horizons.

\begin{center}
\begin{figure}[ht]
\hspace*{-1cm}\includegraphics[scale=0.25]{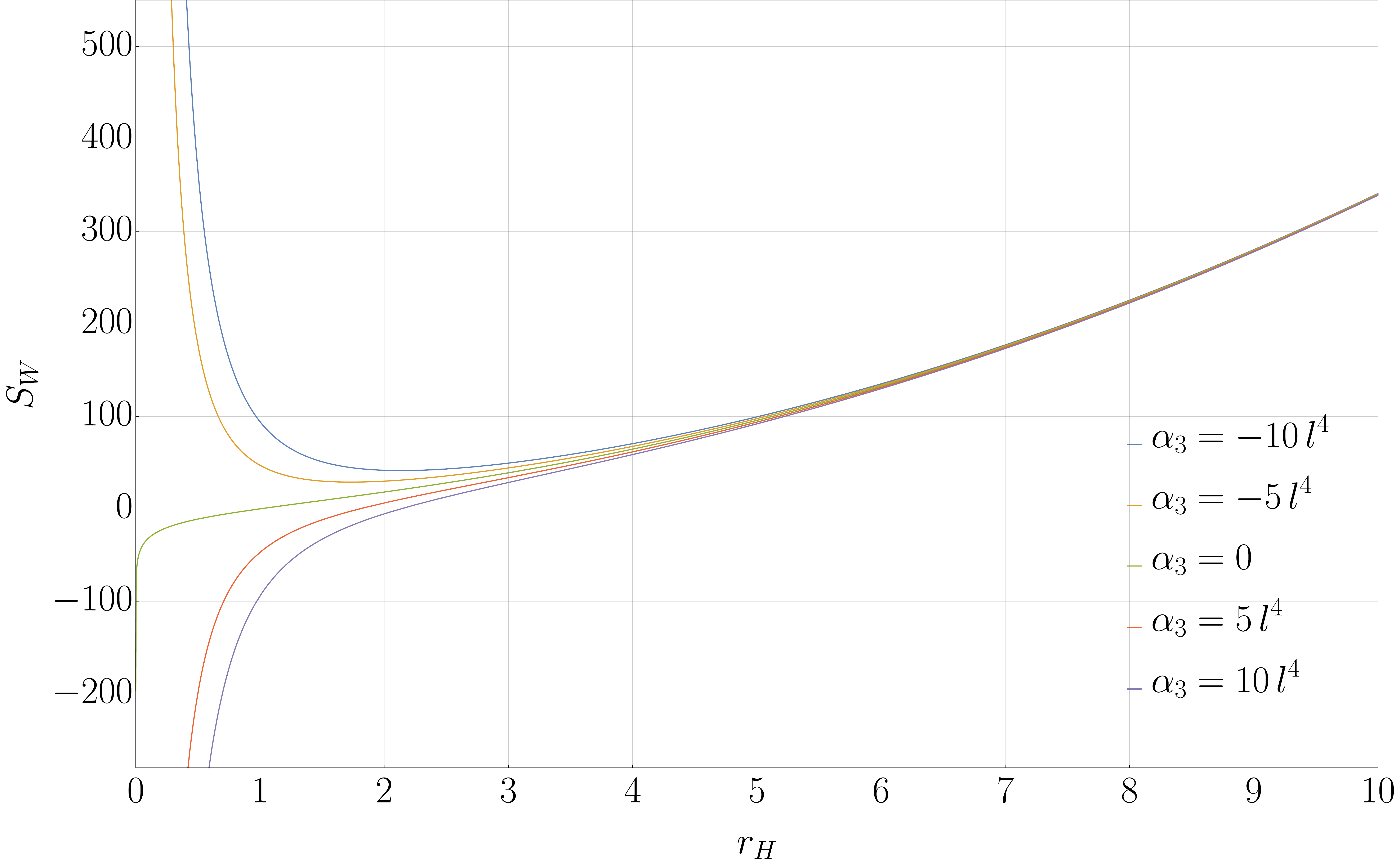}
\caption{Plot of the Wald entropy \eqref{eq:wald_entropy_2}, as a function of the horizon radius $r_H$, in the case $t=3$, with $\alpha_2 = l^2$ and different values of $\alpha_3$.\label{fig:S_w}}
\end{figure}
\end{center}

Concerning its Wald entropy, we plot in Fig. \ref{fig:S_w} the general case for $\alpha \neq \beta$. Among the curves in the plot, consider the ones associated with theories satisfying $\beta=\alpha_3 >0$.
As we can see, the entropy vanishes at some small positive values of $r_H$ in this case. Indeed, for the solution (\ref{eq:A3}) we have $S_W\left(r_H= \sqrt{\alpha}/2\right)>0$, while $S_W\left(r_H= \sqrt{\alpha}\right)<0$. This is also the regime at which the black hole can be extremal, so that a vanishing  entropy can be expected: the extremal regime is determined by $M'(r_H)=0$, so that neglecting for now the cosmological constant we obtain that the horizon radius of the extremal black hole is given $r_H=\sqrt{\alpha} \sqrt{\frac{1+\sqrt{5}}{2}}$, proportional to the square-root of the golden ratio. 

Focusing again on the solution \eqref{eq:A3}, we note that is less singular at the (static) center with respect to the solution in Eq. \eqref{eq:a2r}. In fact, in the first case the metric function $A$ behaves as
\begin{equation}
A(r \to 0)= 1 -r \, \left( \frac{6 M}{\alpha^2}\right)^{1/3} + \frac{r^2}{\alpha} + O\left( r^3 \right),
\end{equation}
which contains a linear correction in the radius $r$, instead of a $\sqrt{r}$ term coming  the Gauss-Bonnet contribution alone. Therefore, it is interesting to investigate what happens for a large number $t-1$ of curvature corrections. 

\begin{center}
\begin{figure}[!htp]
\hspace*{-1cm}\includegraphics[scale=0.23]{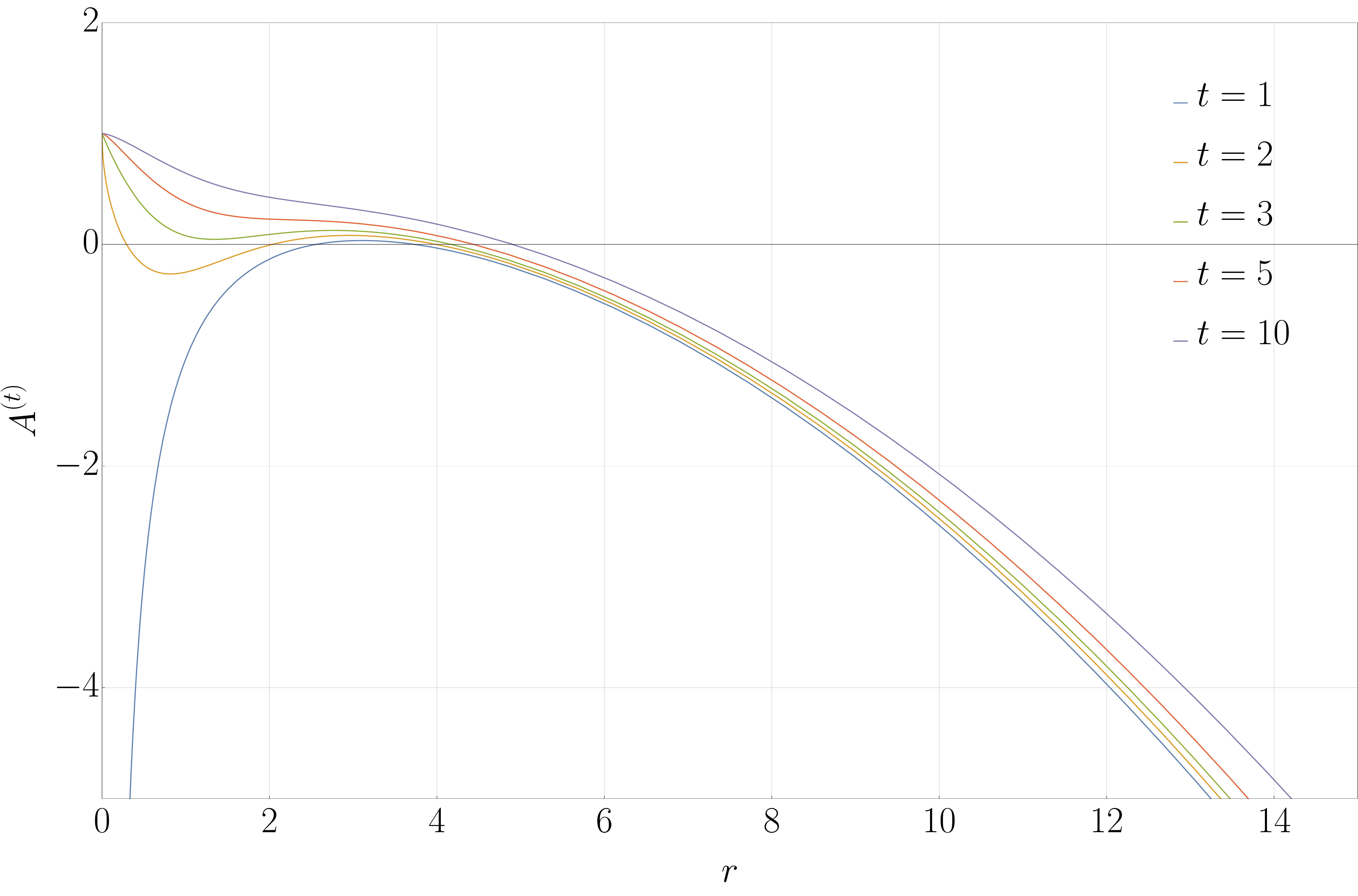}
\caption{Plot of the black hole solution \eqref{eq:At}, as a function of the radius $r$, for different values of the Lovelock order parameter $t$. In this plot we set $k=1$, $\alpha=1$, $M=1$, $\Lambda_0=0.1$.  \label{fig:At}}
\end{figure}
\end{center} 

In order to do so, consider the space of solutions $\left\{ A^{(t)} \right\}_{t>0}$, which contains the black hole solutions up to an arbitrary truncation $t$. The question is whether there are trajectories $A^{(t)}$ (given by a set of coupling constants $\alpha_t$) such that $A^{(1)}$ is the Schwarzschild de-Sitter solution of GR, and $A^{(t \to \infty)}$ is a regular solution. A straightforward generalization of the previous black hole solutions yields to
\begin{equation}
A^{(t)}(r) = k - \frac{r^2}{\gamma} \left[ -1 +   \sqrt[t]{  1+ t\, \gamma \left( \frac{2M}{r^3}  + \frac{\Lambda_0}{3} \right) }\right]\,,\label{eq:At}
\end{equation}
where $\gamma=2^{3-t}\,\alpha$ and the coupling constants are choosen to be
\begin{equation}
    \tilde{\alpha}_p=\frac{\gamma^{p-1}}{t}\binom{t}{p}\,.\label{eq:coup_sss}
\end{equation}
This class of solutions indeed approaches regularity (in particular flatness) as its curvature invariants vanish for $t \to \infty$. For example, neglecting the cosmological constant, the Ricci scalar for $t\to \infty$ is given by 
\begin{equation}
R= \frac{3}{\gamma t} \left(-7+4\log\left[\frac{2 M t \gamma}{r^3}\right] \right) \to 0.
\end{equation}
A plot of the black hole solutions in Eq. \eqref{eq:At} is shown in Fig. \ref{fig:At} for different values of $t$. We also plot the mass function in Fig. \ref{fig:Mt}. As one can notice, the horizons structure is very similar for all the cases plotted with $t>1$.

Moreover, we note that with the choice of the couplings in Eq. \eqref{eq:coup_sss}, in the case $\Lambda_0=0$, we select theories with a unique vacuum, similarly to Lovelock-Born-Infeld and Chern-Simons gravities in higher dimensions, see for example \cite{Lovelock-Born-Infeld1,Lovelock-Born-Infeld2, Lovelock-Chern-Simons}. On the contrary, for $\Lambda_0>0$, the vacuum is dS$_4$, with an effective cosmological constant
\begin{equation}
\Lambda_{\text{eff}} = \frac{-1 + \left(1 +\frac{ t \gamma \Lambda_0}{3}\right)^{1/t}}{\gamma} \,,
\end{equation}
which approaches the zero value as the number of curvature corrections increases, while reducing to $\Lambda_{\text{eff}} = \Lambda_0 /3$ for $t=1$ or $\gamma \to 0$.

The last step is to find the form of the action with the choice of couplings in Eq. \eqref{eq:coup_sss}. The action of these theories is of the form $\sum \alpha_p \mathcal{R}^p = \sum \tilde{\alpha}_p \tilde{\mathcal{R}}^p$, where $\tilde{\mathcal{R}}$ contains the dimensional poles coming from the use of $\tilde{\alpha}$ and we drop the indices for simplicity. Thus we get that the contributions to the action are given by
\begin{equation}
\mathcal{L}^{(t)} \approx -2\Lambda_0 + \frac{-1 + \left(1 + \gamma \tilde{\mathcal{R}}\right)^t}{t \gamma}\,.
\end{equation}

Another way to understand the internal structure of the black hole solutions is to consider the regime where the term of largest order in the curvature series (multiplied by $\alpha_t$) dominates over the low energy ones, i.e. $\alpha_p \mathcal{L}_p  \ll \alpha_t \mathcal{L}_t$ for $p<t$, so that a single term remains in the Lovelock series, which becomes a pure Lovelock theory, see for example \cite{PureLovelock}.
In that case the black hole solution is given by 
\begin{equation}
 ds^2 = - \left( k - \frac{r^2}{l^2} \sqrt[t]{\frac{2 M l^2 }{ r^3}} \right) dt^2 + \frac{1}{k -\frac{r^2}{l^2}   \sqrt[t]{\frac{2 M l^2 }{ r^3}}} dr^2 + r^2 d\Omega_{2}^2\,,
\end{equation}
where we set $\tilde{\alpha}_t = l^{2(t-1)}$, with $l$ a very small length scale with respect to the other neglected coupling constants.
Therefore, the solution becomes less and less singular as the order of the Lovelock term $t \to \infty$ because it approaches a de Sitter solution. For example, its Ricci scalar is given by  
\begin{equation}
R= \frac{3(t-1)(4t-3)}{ l^2 t^2} \left(\frac{2 M l^2}{r^3}\right)^{1/t} \to \frac{12}{l^2}.
\end{equation}
 Concerning the existence of non-singular black hole and cosmological solutions in the non-perturbative case ($t\to \infty$), see \cite{aim}.

\section{Conclusions}\label{sec:concl}

In this paper we have explored specific sectors of Lovelock-Lanczos gravity of generic order $p$ by using a non-standard prescription, which makes it non-trivial at and beyond the critical order $d\leq 2p$, for any dimensions $d>2$. This leads to almost dimensional independent results, and allows to find well-defined results for FLRW and spherically symmetric spacetimes, as well as at first order in perturbation theory on (anti-)de Sitter backgrounds.

In the framework of cosmological solutions, the main difference between this theory and standard GR is the existence of one or more de Sitter vacua. This fact is important for the phenomenology of both Late and Early Universe. In the former case, a quite generic stable de Sitter attractor is a desirable feature since it avoids the introduction of a fine-tuned cosmological constant. Another very interesting feature of this model is that the perturbation equations are the same as in GR, at least at the first order. This is of particular relevance since it indicates that the addition of Lovelock terms modifies the background evolution but, maybe, leaves untouched the standard evolution of large structures. 

Concerning the black hole solutions of this theory, we saw that the curvature singularities can be cured by considering the non-perturbative limit ($t\to\infty$). 
Together with the fact that the first correction yields a logarithmic entropy, this might indicate that this regularized theory could constitute a suitable effective action for gravity. In this spirit, it would be interesting to compare our results with the so-called (Generalized)-Quasi-Topological gravity models, which were found to constitute a very general effective (metric) theory of gravity \cite{GQTEffAct1,GQTEffAct2}. As these theories share many features with the Lovelock-Lanczos models, in particular the possibility to be regularized\footnote{Indeed, it can be checked that the limit $d\to4$ of the theory given in Eq(2.9) of \cite{1703.01631} and Eq(3.10) of \cite{Quasi-Top1}, which has a pole $1/(d-4)$, can be taken after evaluating it to (dynamical) spherical symmetry, see Eq(30) of \cite{aim}.}, it might be interesting to consider both at the same time and reiterate some investigations about arbitrary order black holes \cite{arborderBH}, as well as geometric inflation and regular cosmological solutions \cite{geomInfl}.

Moreover, as we saw, these kinds of theories can have more than one vacuum, so that they can provide a natural high energy AdS$_4$ vacuum, in addition to the usual dS$_4$ one of GR with a cosmological constant. Thus, this branch of the theory's dynamics might be possible to investigate with the tool of the AdS/CFT correspondence. Finally, the precise relation between the regularized theory admitting the family of black hole solutions Eq(\ref{eq:At}) with a unique vacuum and Lovelock-Born-Infeld or Lovelock-Chern-Simons gravities  \cite{Lovelock-Born-Infeld1, Lovelock-Born-Infeld2, Lovelock-Chern-Simons}  should be investigated further.

We stress that the number of classes of metric fields for which this prescription works is still quite small, and it would be very interesting to look for similar regularization procedures of Lovelock gravity for rotating black holes or anisotropic cosmological backgrounds, in order to understand better the possible measurable effects induced by these terms. Some very recent results along these directions are discussed in \cite{DALLG}.

Finally, as we mentioned in the introduction, there are many well-defined four dimensional theories admitting the ``minisuperspace regularized" sectors of Gauss-Bonnet gravity that we have generalized here. In order for our results to be fully applicable in four dimensions, it is thus necessary to find if such property is preserved for the regularizations of higher order Lovelock-Lanczos gravity. This would automatically yield to four-dimensional ``Lovelock"-like theories with regular solutions and might thus deserve further investigations.

\subsection*{Acknowledgments}

We thank Prof.\ A.\ Cisterna for useful discussion about this work, Peng-Cheng Li for pointing out to us the paper \cite{conformal} and Chunshan Lin for the paper \cite{GRD=2}. This work has been partially performed using the software \texttt{xAct} \cite{xAct}. A.\ C.\ acknowledges the financial support of the Italian Ministry of Instruction, University and Research (MIUR) for his Doctoral studies. S.\ V.\ acknowledges the financial support of the Italian National Institute for Nuclear Physics (INFN) for her Doctoral studies. Finally, we thank the referees for their suggestions and the opportunity to improve our paper.

\newpage{\pagestyle{empty}\cleardoublepage}

\end{document}